\documentclass[prd,aps,floats,floatfix,superscriptaddress,preprintnumbers,
showpacs,eqsecnum,nofootinbib,twocolumn]{revtex4}
\usepackage{latexsym,array,theorem,mathrsfs,subfigure,bm,float}
\usepackage{latexsym,array,theorem,mathrsfs,subfigure,bm,float}
\usepackage{psfrag}
\usepackage{amsfonts,amsmath,amssymb,latexsym,array,afterpage,theorem,mathrsfs,bm,float,epsfig,color,graphicx,tabularx,here,multirow,cases}

\newcommand{\nn}{\nonumber \\}
\newcommand{\bea}{\begin{eqnarray}}
\newcommand{\ena}{\end{eqnarray}}
\newcommand{\beann}{\begin{eqnarray*}}
\newcommand{\enann}{\end{eqnarray*}}

\begin{document}

\baselineskip=12pt

\newcommand{\KM}{\color{red}}

\newcommand{\MF}{\color{blue}}

\newcommand{\MFg}{\color{green}}

\title{Gravitational Baryogenesis after Anisotropic Inflation}
\author{Mitsuhiro \sc{Fukushima}}
\email{dark-matter@gravity.phys.waseda.ac.jp}
\affiliation{
Department of Physics, Waseda University,
3-4-1 Okubo, Shinjuku, Tokyo 169-8555, Japan
}

\author{Shuntaro \sc{Mizuno}}
\email{shuntaro.mizuno@aoni.waseda.jp}
\affiliation{Waseda Institute for Advanced Study, Waseda University, 
1-6-1 Nishi-Waseda, Shinjuku, Tokyo 169-8050, Japan
}

\author{Kei-ichi \sc{Maeda}}
\email{maeda@waseda.ac.jp}
\affiliation{
Department of Physics, Waseda University,
3-4-1 Okubo, Shinjuku, Tokyo 169-8555, Japan
}

\date{\today}

\begin{abstract}

The gravitational baryogensis may 
not generate a sufficient baryon asymmetry in the standard thermal history of the Universe  
when we take into account the gravitino problem.
Hence it has been suggested that  anisotropy of the Universe can enhance
the generation of the baryon asymmetry through 
the increase of the time change of the Ricci scalar curvature.
We study the gravitational baryogenesis  
in the presence of anisotropy, which is produced at the end of 
an anisotropic inflation.
Although we confirm that the generated baryon asymmetry is 
enhanced compared with the original isotropic cosmological model,
taking into account the constraint on the anisotropy by the recent CMB observations, 
we find that it is still difficult to 
obtain the observed baryon asymmetry only through the gravitational baryogenesis without suffering from the gravitino problem.

\end{abstract}


\pacs{98.80.-k, 98.80.Cq}

\maketitle

\section{Introduction}

The thermal history of the Universe after the Big Bang Nucleosynthesis (BBN) is well studied, and some observations have confirmed it as the standard cosmology.
However, it suffers from various initial conditions such as the horizon problem, flatness problem, relic problem such as monopoles and the origin of the large scale structure of the Universe.
Fortunately, these problems are elegantly solved all at one time  by the idea of 
inflation \cite{Inf}.
Nevertheless the standard Big Bang cosmology still have some 
unsolved problems. 
One is the baryon-antibaryon asymmetry appeared before the BBN epoch.
Its origin has been remained as an outstanding mystery of the particle cosmology.
In the inflationary scenario, 
as any pre-existing baryon asymmetry would be rapidly diluted, 
the baryon asymmetry must be generated after inflation.

The baryon asymmetry is typically characterized by the ratio of the baryon number density $n_B$ to the entropy density $s$.
Its value is observationally obtained from the highly precise measurement 
of the cosmic microwave background radiation (CMB) by {\it Planck} mission \cite{Planck} as
\begin{equation}
Y_B\equiv\frac{n_B}{s}=\left(0.864^{+0.016}_{-0.015}\right)\times10^{-10}. \label{ObsYb}
\end{equation}
The other independent observations such as the abundance of the primordial light elements from BBN also support this value \cite{BBN}.
In order to generate a non-vanishing $Y_B$, Sakharov \cite{Sakharov} has 
argued the following three necessary conditions: (i) the existence of baryon number ($B$) violating interactions; (ii) the breaking of 
$C$ and $CP$ symmetries; and (iii) departure from thermal equilibrium.
Many baryogenesis models which satisfy the above criteria have so far been proposed \cite{GUTBG, EWBG, LG, ADBG}.

Interestingly, however, there exists some loopholes to generate the baryon asymmetry 
without satisfying the Sakharov criteria.
For example,  in Ref.~\cite{SBG},  the effective $CPT$ violating interaction is introduced\footnote{The recent studies of spontaneous baryogenesis and its extensions are as follows \cite{SBG2, SBG3}.}.
Since the $CPT$ violating interaction can bias $B$-violating interaction among particles and antiparticles in the thermal equilibrium, the Sakharov's third criterion is not required with this interaction.
As the similar idea,
the possibility that a gravitational interaction plays 
an interesting role in baryogenesis has been proposed in {Ref.}~\cite{GBG}.
It was shown that such an interaction dynamically breaks $CPT$ in an expanding Universe 
and generates the baryon number asymmetry
while maintaining thermal equilibrium.

This gravitational baryogenesis is one of 
the attractive models as its interaction may be obtained in supergravity theories.
However, in order to generate observationally sufficient baryon asymmetry  
(\ref{ObsYb}), the Universe needs to experience a high-enough temperature state during a generating baryon asymmetry phase if the Universe follows the standard thermal history.
Unfortunately this condition conflicts with the requirement from the gravitino problem not to overproduce the lightest supersymmetric particles (LSPs) \cite{GProb, ReheatTemp}.
Hence, in the paradigm of supergravity, the gravitational baryogenesis seems not 
to work well.

In a less symmetric background spacetime, however,
some possibility of the enhancement of the baryon asymmetry 
was argued in Ref.~\cite{A-GBG}.
They changed the background from the Friedmann-Lema\^itre-Robertson-Walker (FLRW) 
isotropic universe to the Bianchi type I anisotropic spacetime and showed that the baryon asymmetry will increase with the anisotropy of the Universe.
However, they have not discussed the detail, i.e., 
how large  this anisotropic enhancement effect is and whether or not it can 
really solve the aforementioned problem.
In addition, they have not mentioned about the origin of anisotropy at all, either.

Hence, in this paper, we consider the Universe which has an anisotropic characteristic at the early stage of its history and assume that this anisotropy originates from an anisotropic inflationary model \cite{AInf}.
Although most inflationary scenarios assume an isotropic spacetime,
there is a possibility that an inflation may be influenced by the existing gauge field
 which coupled to the inflaton field \cite{Gauge-Inf}.
Although, in the first place, these models are motivated by generating primordial magnetic field, they also generate a statistical anisotropy of the curvature perturbations \cite{AnisoP}.
Of course, as long as we accept the cosmic no-hair conjecture \cite{CNoH, CNoH2, CNoH3}, 
the accelerated expansion during inflation makes the Universe isotropic.
However, we can evade this conjecture by introduction of
 nonminimal kinetic term of a vector field inspired from the supergravity theory.
An anisotropic hair can survive during and after the inflation.
We find that the anisotropy of the Universe rapidly grows at the end of the inflation, 
which may affect the gravitational baryogenesis.
We evaluate how large anisotropy is generated after the anisotropic inflation 
and find how large enhancement of the gravitational baryon asymmetry 
is achieved.
Then we conclude whether or not the gravitational baryogenesis can explain 
the observed baryon asymmetry in the context of the anisotropic inflation.

The rest of this paper is organizes as follows.
In Sec.~\ref{GBGandEx}, we will overview the gravitational baryogenesis model 
in the standard Big Bang cosmology and 
explain how it is constrained by the gravitino problem.
In Sec.~\ref{Gravitational_Baryogenesis_Anisotropic_Inflation}, 
we study the gravitational baryogenesis in the anisotropic inflationary scenario 
and  evaluate how the amount of baryon asymmetry is enhanced by 
the anisotropy of the Universe.
The last section is devoted to discussion and conclusions.
In Appendix, we shortly summarize the anisotropic inflation.

\section{Gravitational Baryogensis} \label{GBGandEx}

\subsection{Gravitational Baryogenesis}
In order to discuss gravitational baryogenesis, we consider  the following interaction:
\begin{equation}
\mathcal{S}_{\text{int}} = \frac{1}{M_\ast^2}\int d^4x\sqrt{-g}\left(\partial_\mu R\right)J^\mu \label{eqGBG},
\end{equation}
where  $R$ is the Ricci scalar curvature and $J^\mu$ is the baryon number current
($J^\mu$ could be the $B-L$ charge current, where $B$ and $L$ stand for baryon and lepton number, respectively, 
and  $B-L$ can be translated to $B$ via electroweak sphaleron process {\cite{Sphaleron}.).
$M_\ast$ is a cut-off mass parameter in an effective theory.
If  $M_\ast$ is of the order of magnitude of the reduced Planck mass $M_P \equiv (8\pi G)^{-1/2}\simeq 2.4\times10^{18}$ GeV, 
the above interaction could be obtained in a low-energy effective field theory of quantum gravity. 
It is also worth mentioning that
this interaction can be obtained from a higher dimensional operator in the K$\ddot{\rm a}$hler potential 
in supergravity theories \cite{GBG}. 

Since the interaction in Eq.~(\ref{eqGBG}) violates $CP$, 
if there exists a $B$-violating process in thermal equilibrium,
it can generate a net baryon number.
The generated net baryon number can be evaluated as follows:
In the expanding homogeneous Universe,
 we have
\begin{equation}
\frac{1}{M_\ast^2}\left(\partial_\mu R\right)J^\mu = \frac{\dot{R}}{M_\ast^2}\left(g_bn_b+g_{\bar{b}}n_{\bar{b}}\right),
\end{equation}
where $g_b=-g_{\bar{b}}\sim\mathcal{O}(1)$ are the baryon number, $n_b$ and $n_{\bar{b}}$ stand for the number densities of baryon and anti-baryon, respectively.
In what follows,
we will use an overdot sign to
denote the derivative with respect to the cosmic time.
This interaction shifts the energy of a baryonic particle
 by the amount of $2g_b\dot{R}/M_\ast^2$ relative to that of an anti-baryonic particle,
 which provides an effective ``chemical'' potential given by
$\mu_{b} = g_b\dot{R}/M_\ast^2 = -\mu_{\bar{b}}$.

For relativistic baryons,
 using this effect,
in thermal equilibrium, 
the non-zero baryon number density given by
\begin{equation}
n_B = g_b(n_{b} - n_{\bar{b}}) \simeq -\frac{g_b\mu_b}{6}T^2\,,
\end{equation}
will be generated \cite{EarlyUniverse}.
Here, we have assumed $T\gg|\mu_b|$.

In the expanding Universe, when the temperature drops and $B$-violating interactions become ineffective, a non-zero value of $n_B$ will be frozen.
Therefore, the net baryon asymmetry remains
below the decoupling temperature $T_D$, where 
we denote the epoch when $B$-violating interactions is 
frozen out by the subscript $D$. 

While the entropy density of the Universe is given by $s=2\pi^2g_\ast(T) T^3/45$
where $g_\ast(T)$ denotes the total degree of freedom for particles that contribute to the entropy of the Universe.
Consequently, the baryon asymmetry parameter is given by
\begin{equation}
Y_B \equiv \frac{n_B}{s} \simeq -\left.\frac{15g_b^2}{4\pi^2g_\ast(T)}\frac{\dot{R}(T)}{M_\ast^2T}\right|_{T_D} \,. \label{eqYB}
\end{equation}

For the spatially flat FLRW Universe, 
whose metric is given by
\begin{equation}
ds^2 = -dt^2 +a^2(t)\left(dx^2+dy^2+dz^2\right),
\end{equation}
the Einstein equations become
\begin{equation}
H^2 = \frac{1}{3M_P^2}\rho \,, ~~
\dot{H} = -\frac{1}{2M_P^2} (P+\rho)\,, \label{IsoEin}
\end{equation}
where $H$ is the expansion  Hubble parameter defined by $\dot{a}/a$, 
$P$ and $\rho$ are the pressure and energy density of 
matter fluid, respectively.
The  effective equation-of-state (EOS) parameter $w=P/\rho$ is not necessary 
to be constant.

In order to fix the decoupling temperature $T_D$,
 we have to specify the $B$-violating interaction.
 In this paper, we assume  one $B$-violating interaction, which is 
is given by an operator $\mathcal{O}_B$ of mass dimension $4+n$.
Such a $B$-violating interaction may exist as a non-renormalizable interaction 
in some effective field theory 
when we regard the baryon conservation as an eventual symmetry of 
the Standard Model of particle physics. 
$n>0$ is required for the $B$-violating interaction.

Since the coupling constant is proportional to $1/M_B^n$ in this interaction,
where $M_B$ is the mass scale associated with $\mathcal{O}_B$, 
the generation rate of such an interaction in thermal equilibrium with the temperature 
$T$ is given by \cite{GBG}
\begin{equation}
\Gamma_B = \frac{T^{2n+1}}{M_B^{2n}}\,.
\label{B-violate}
\end{equation}
The $B$-violating interaction is decoupled when the Hubble parameter becomes 
larger than $\Gamma_B$.
Therefore, $T_D$ is fixed by the condition $H=\Gamma_B$.
$T_D$ is described by some function of $M_B$ and $n$,
 which are most fundamental parameters in the $B$-violating interaction model.
 It is worth mentioning that a decoupling of $B$-violating dimension 5 interaction
  does not occur during reheating phase.
This is because the $B$-violating interaction rate $\Gamma_B$ with the mass dimension
smaller than 6 decreases always slower than the Hubble parameter $H$, and 
then $H$ cannot exceed $\Gamma_B$.

When we discuss the gravitational baryogenesis, we have to evaluate 
the Ricci scalar curvature $R$, which is given by
\begin{equation}
R = 3H^2(1-3w).
\end{equation}
Using the Einstein equations (\ref{IsoEin}),  the time derivative of the Ricci scaler curvature
 is written by
\begin{equation}
\dot{R} = -\sqrt{3}(1+w)(1-3w)\frac{\rho^{3/2}}{M_P^3}-3\dot{w}\frac{\rho}{M_P^2}. \label{eqdotR}
\end{equation}

In order to evaluate the generated net baryon asymmetry,  
we need to specify the epoch of $B$-violation decoupling and calculate the value of $\rho$ and $T$ at that time.
In what follows, 
we will consider just the following two cases: the reheating phase $(w\approx 0)$ and radiation-dominated phase 
$(w\approx 1/3)$
\footnote{It is worth mentioning that  it was reported in Ref.~\cite{GBG} that
 the gravitational baryogenesis works in a very efficient way and easy to explain
the observed baryon asymmetry when the baryon generation occurs in the phase dominated 
by a nonthermal component with
$w>1/3$. This case also includes the other scenarios such as \cite{ekpyrotic, cyclic, quintessence}}.

\subsubsection{Reheating Phase}

The evolution of the thermal plasma during reheating phase is shown in Ref.~\cite{EarlyUniverse}.
If the reheating process after inflation is due to the decay of an inflaton scalar field  $\phi$
 to relativistic particles (radiation) and it is characterized by the oscillation of $\phi$, 
the reheating period can be described by matter-dominated era ($w\approx 0$).
In this phase, the inflaton and radiation energy density evolve as
\bea
&&\dot{\rho}_\phi +3H\rho_\phi = -\Gamma_\phi \rho_\phi ,\\
&&\dot{\rho}_r +4H\rho_r = \Gamma_\phi \rho_\phi ,
\ena
where $\Gamma_\phi$ is the decay rate of inflaton into radiation.

If $M_I^4$ is the vacuum energy of the inflaton field at the end of inflation,
 the energy density of inflaton field is given by
\begin{equation}
\rho_\phi = M_I^4 \left(\frac{a}{a_{\text{osc}}}\right)^{-3} e^{-\Gamma_\phi(t-t_{\text{osc}})}
\,,
\end{equation}
where $t_{\text{osc}}\simeq M_P/M_I^2$.
During this phase, we have $\rho_\phi\propto a^{-3}$ and $a\propto t^{2/3}$.
Using the second law of thermodynamics and supposing that all of the 
energy released from inflaton decay is rapidly converted into radiation, 
we can also evaluate the entropy and radiation energy density in this phase:
\beann
&&S^{4/3} =\frac{4}{3}\left(\frac{2\pi^2g_\ast}{45}\right)^{1/3} \Gamma_\phi M_I^4 
a_{\text{osc}}^4\int_{t_{\text{osc}}}^t \frac{a}{a_{\text{osc}}}e^{-\Gamma_\phi(t-t_{\text{osc}})} dt , \\
&&\rho_r = \Gamma_\phi M_I^4 \left(\frac{a}{a_{\text{osc}}}\right)^{-4}\int_{t_{\text{osc}}}^t 
\frac{a}{a_{\text{osc}}}e^{-\Gamma_\phi(t-t_{\text{osc}})} dt ,\label{REDre}
\enann
where we have ignored the initial entropy.
Since $a\propto t^{2/3}$ in this epoch, we find that
$\rho_r\propto a^{-3/2}$ and $S\propto a^{15/8}$
 if the reheating process is slow $(\Gamma_\phi\ll H(t_{\text{osc}}))$.
Then
\bea
&&\rho_\phi = \frac{\pi^2 g_\ast}{30}T_{RD}^4\left(\frac{a_{RD}}{a}\right)^3 \,,\\
&&\rho_r = \frac{\pi^2 g_\ast}{30}T_{RD}^4\left(\frac{a_{RD}}{a}\right)^{3/2} \,,
\label{eq_reh_rad}
\ena
where the variables with the subscript $RD$ denote those evaluated when radiation becomes 
dominant.  $T_{RD}$ corresponds to the reheating temperature.
Finally, we rewrite the energy density of inflaton field:
\begin{equation}
\rho_\phi = \frac{\pi^2 g_\ast}{30}\frac{T^8}{T_{RD}^4}.
\label{rho_phi}
\end{equation}
The total energy density $\rho$ is approximated by $\rho_\phi$, 
since the main component of the matter field in this phase is the inflaton field.

Using the above result together with
Eq.~(\ref{eqYB}) and Eq.~(\ref{eqdotR}), we evaluate the baryon asymmetry as
\begin{equation}
Y_B \simeq \frac{\pi g_b^2 g_\ast^{1/2}}{8\sqrt{10}} \frac{T_D^{11}}{M_\ast^2 M_P^3 T_{RD}^6}. \label{YbIre}
\end{equation}
This asymmetry, however, is diluted by a continuous production of entropy during the reheating epoch.
From the above, the entropy is generated as $S\propto a^{15/8}$ which means that
the entropy density evolves as $s=S/a^3 \propto a^{-9/8}$.
During reheating phase, the baryon asymmetry dilutes as $Y_B \propto a^{-15/8}$.
In terms of the temperature, this dilution factor is given by
$(T_{RD}/T_{D})^5$, which yields
\begin{equation}
Y_B \simeq \frac{\pi g_b^2 g_\ast^{1/2}}{8\sqrt{10}} \frac{T_D^{6}}{M_\ast^2 M_P^3 T_{RD}}. \label{YbIre2}
\end{equation}
It is worth mentioning that the overall factor in this formula is of order unity since $g_\ast\sim\mathcal{O}(10^2)$ in this time.

In this phase, by definition, the decoupling temperature must be in the range of $T_{RD}<T_D<M_I$.
Meanwhile,  since the observed baryon asymmetry is about $10^{-10}$ 
and  it is natural to assume that the baryon asymmetry before dilution is less than order unity, 
we find the upper bound of the decoupling temperature as $T_{D} \lesssim 10^{2} T_{RD}$.
In consequence, the possible range of the decoupling temperature is
\begin{equation}
T_{RD} < T_D \lesssim 10^2 T_{RD}.
\end{equation}

On the other hand, $T_D$ needs to be expressed by using $M_B$ that is a typical mass scale of the $B$-violating interaction.
From Eq.~(\ref{IsoEin}) and Eq. (\ref{rho_phi}), the Hubble parameter at the decoupling time is evaluated by
\begin{equation}
H(t_D) = H(t_{RD})\frac{t_{RD}}{t_D} = \frac{\pi g_\ast^{1/2}}{3\sqrt{10}}\frac{T_D^4}{M_PT_{RD}^2},
\end{equation}
where we have used $t\propto T^{-4}$ in this epoch.
Using Eq.~(\ref{B-violate}), the decoupling temperature is fixed as
\begin{equation}
T_D \simeq \left(\frac{M_B^{2n}}{M_PT_{RD}^2}\right)^{1/(2n-3)}\,. \label{TdRe}
\end{equation}
As a result, the final baryon asymmetry (\ref{YbIre2}) is rewritten as
\begin{equation}
Y_B \simeq M_B^{\frac{12n}{2n-3}}M_\ast^{-2}T_{RD}^{-\frac{2n+9}{2n-3}}M_P^{-\frac{6n-3}{2n-3}}.
\end{equation}

The range of $M_\ast$ and $M_B$, in which  the observationally acceptable baryon asymmetry
 is found,
is shown in Fig.~\ref{fig01},
assuming a dimension-6 $B$-violating interaction ($n=2$).

\begin{figure}[tbp]
 \centering
 \includegraphics[width=8cm,clip]{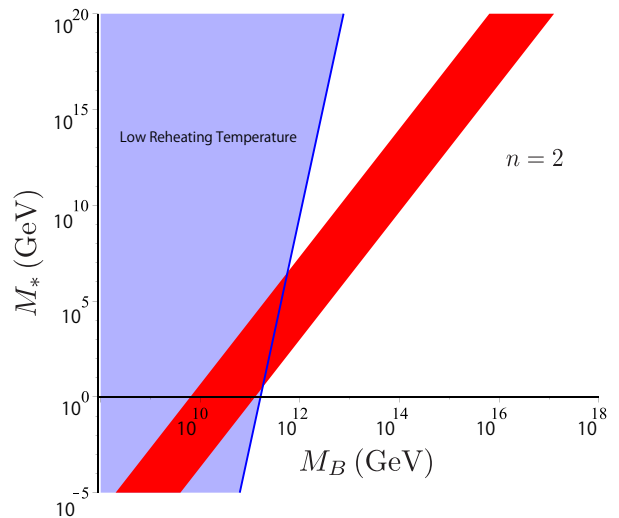}
 \caption{
The acceptable range of parameters $M_\ast$ and $M_B$ in
the case that the decoupling occurs during the reheating phase.
We assume a dimension-6 $B$-violating interaction ($n=2$).
The red region is required  to explain  the observed baryon asymmetry  ($Y_B\sim10^{-10}$).
The lower and upper bounds correspond to  $T_D=T_{RD}$ and  $T_D=10^2 T_{RD}$, respectively. 
In supersymmetric theories, however,  we have an additional 
constraint on the reheating temperature 
by gravitino problem ($T_{RD} < 10^9$ GeV), 
whose acceptable parameter range  is shown by the blue region.
The intersection of the red and blue regions may explain baryon number asymmetry.
}
 \label{fig01} 
\end{figure}

\subsubsection{Radiation-dominated Phase}

Next, we check the case of the radiation-dominated era after the reheating phase.
Although this case is simpler than the previous one, 
we have to take into account 
some non-trivial effects.
The radiation-dominated epoch 
is characterized by $w=1/3$ and then $R=0$.
As a result, no net baryon asymmetry seems to be generated.
However, we find a loophole for this issue by taking into account the quantum anomaly effect \cite{TraceAnomaly}
\footnote{We also find that in the modified gravity theory, e.g.,   in $f(R)$ gravity, 
a net baryon asymmetry may be generated even during the radiation dominated era
\cite{ModGR-GBG}.}.
The typical gauge fields and matter contents at very high energy scale 
have a trace anomaly, whose equation of state (EOS) is given by  $(1-3w\sim10^{-2}\text{-}10^{-1})$.
This trace anomaly makes $T^\mu_\mu\neq0$ and can generate a net baryon asymmetry.
In what follows, we discuss this case.

Using the above anomaly effect,
we can evaluate the amount of the generating baryon asymmetry as
\begin{equation}
Y_B \simeq \frac{\pi g_b^2 g_\ast^{1/2}}{6\sqrt{10}}(1-3w) \frac{T_D^5}{M_\ast^2 M_P^3} , 
\label{YbIRD}
\end{equation}
where  we have used 
\begin{equation}
\rho_r = \frac{\pi^2g_\ast}{30}T^4. \label{ERad}
\end{equation}
The overall factor of $Y_B$ is approximated by $(1-3w)$.
In this case, the decoupling temperature must satisfy the condition
\begin{equation}
T_D < T_{RD} < M_I .
\end{equation}

In the similar manner to the previous reheating case, we rewrite $T_D$ in terms of $M_B$.
The Hubble parameter at the decoupling time is given by
\begin{equation}
H(t_D) = \frac{\pi g_\ast^{1/2}}{3\sqrt{10}}\frac{T_D^2}{M_P}
\,,
\end{equation}
and then using Eq.~(\ref{B-violate}), the decoupling temperature is written by
\begin{equation}
T_D \simeq \left(\frac{M_B^{2n}}{M_P}\right)^{1/(2n-1)}
\,.\label{TdRD}
\end{equation}
Hence, the baryon asymmetry given by Eq.~(\ref{YbIRD}) is evaluated as
\begin{equation}
Y_B \simeq (1-3w)M_B^{\frac{10n}{2n-1}}M_\ast^{-2}M_P^{-\frac{6n+2}{2n-1}}
\,.
\label{YbIRD2}
\end{equation}

In the radiation dominated era, the range of $M_\ast$ and $M_B$, 
in which the observed baryon asymmetry is obtained,  is
shown in Fig.~\ref{fig02}, 
in which  we have assumed a dimension-6 $B$-violation interaction  ($n=2$).

\begin{figure}[tbp]
 \centering
 \includegraphics[width=8cm,clip]{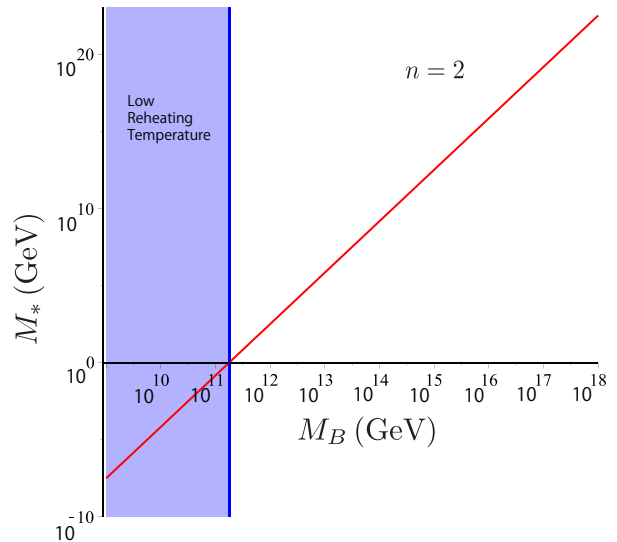}
 \caption{
The acceptable range of parameters $M_\ast$ and $M_B$ in
the case that the decoupling occurs during the radiation-dominated era.
We assume a dimension-6 $B$-violating interaction ($n=2$) and 
 the trace anomaly with $1-3w=10^{-1}$.
The red region is required  to explain  the observed baryon asymmetry  ($Y_B\sim10^{-10}$).
In supersymmetric theories, however,  we have an additional 
constraint on the reheating temperature 
by gravitino problem ($T_{RD} < 10^9$ GeV), 
whose acceptable parameter range  is shown by the blue region.
The intersection of the red and blue regions may explain baryon number asymmetry.
}
 \label{fig02} 
\end{figure}

\subsection{Additional Constraint  in Supergravity Theories}

So far, we have discussed the condition for 
which the observationally acceptable baryon asymmetry can be generated.
However, since some $B$-violating interactions may be expected in supergravity theories,
we will have an additional constraint in such a model. 

In supergravity theories, 
it is well known that gravitino production places severe bounds on $T_{RD}$.
This bound comes from two constraints: (i) protecting the products of BBN from decay which is caused by late gravitino decays, and (ii) avoiding the
overclosure of the Universe by gravitinos.
Since the exact value of  $T_{RD}$
is related with the gravitino mass $m_{3/2}$ and decaying process of gravitino which strongly depend on supergravity theories,
throughout this paper, we adopt the constraint that  $T_{RD}$ must be smaller than $10^9$ GeV, obtained from typical supergravity theories \cite{ReheatTemp}\footnote{This restriction comes from the observational constraints of $^4$He abundance with assuming 10 TeV gravitino mass.}.


From this restriction, using Eqs.~(\ref{TdRe}) or  (\ref{TdRD}), the allowed region of $M_B$ is limited from the above.
Consequently, this makes $M_\ast$ also small in order to explain the observationally acceptable baryon asymmetry.

In Figs.~\ref{fig01} and \ref{fig02}, then we show how the allowed parameter region will be further restricted by this 
additional constraint.
From Fig.~\ref{fig01}, the allowed region predicts a tiny value of $M_\ast$,
 which is lower than LHC energy scale.
Since we have not so far found any $B$-violating interactions up to such a scale,
we conclude that the gravitational baryogenesis during the reheating phase is unfavorable.
Thus, we can conclude that the gravitational baryogenesis during the reheating phase is unfavorable.
From Fig.~\ref{fig02}, 
the cut-off mass scale $M_\ast$ also have to be small in order to fulfill the observational
constraint.
As a result, we may not expect the efficient generation of baryon asymmetry in the FLRW background.
That is why we have to look for some enhancement mechanisms of generating baryon asymmetry
for the  gravitational baryogenesis in supergravity theories.

In next section, we shall consider some enhancement mechanism, 
which is found in an anisotropic universe \cite{A-GBG}, and analyze the detail assuming 
an anisotropic inflationary model.

\section{Gravitational Baryogenesis in Anisotropic Inflation}
\label{Gravitational_Baryogenesis_Anisotropic_Inflation}

\subsection{Anisotropic Extension of Gravitational Baryogenesis}

As mentioned above,  it seems difficult to generate
the baryon asymmetry  by the
gravitational baryogenesis both in the reheating phase and in the radiation dominated 
stage after inflation.
However, the baryon number asymmetry could be enhanced in a
 less symmetric background spacetime
as pointed out in Ref.~\cite{A-GBG}.
They assumed the Bianchi type I anisotropic background spacetime but with
an isotropic matter fluid, 
and then they found that the effect of anisotropy  enhances
the generated baryon asymmetry compared with that in the FLRW background.

Here, we will briefly summarize the mechanism  in Ref.~\cite{A-GBG} and extend their discussion.
\begin{widetext}
We consider an isotropic matter field and  Bianchi type I spacetime with the metric
\begin{equation*}
ds^2 = -dt^2+e^{2\alpha(t)}\left[
e^{-4\beta_+(t)}dx^2+e^{2(\beta_+(t)+\sqrt{3}\beta_-(t))}dy^2
+e^{2(\beta_+(t)-\sqrt{3}\beta_-(t))}dz^2\right]\,.
\end{equation*}
\end{widetext}
Then the time derivative of the scalar curvature, 
 is given by
\begin{equation}
\dot{R} = -\sqrt{3}(1+w)(1-3w)\frac{\rho\sqrt{\rho+3M_P^2\Sigma^2}}{M_P^3}
\,.\label{A-dotR}
\end{equation}
 $w=P/\rho$ is the EOS parameter of the isotropic matter field. 
$\Sigma$ is the magnitude of the shear of anisotropic expansion, which is 
defined by
\begin{equation}
\Sigma^2 = {\dot{\beta}_+}^2+
{\dot{\beta}_-}^2\,,
\end{equation}
The mean expansion rate $H$, which corresponds to 
the Hubble expansion parameter in the FLRW spacetime, is defined by 
\bea
H=\dot \alpha
\,.
\ena

It is worth mentioning that this anisotropic background brings 
two enhancement effects through $\dot{R}$ and $T_D$.
The shear term 
increases  $\dot{R}$ and then it 
enhances the generated baryon asymmetry
because  $Y_B$ is proportional to $\dot{R}$.
In contrast, since the effect through $T_D$ is non-trivial, 
one may need a further explanation, which is given as follows.
In what follows, we focus on the case when the decoupling of the $B$-violating interaction occurs
 during the radiation-dominated phase, for simplicity.
$T_D$ is determined by the condition $H\sim \Gamma_B$. 
Since $H$ and $\Gamma_B$ are determined by 
the Friedmann equation of the anisotropic Universe,
\begin{equation}
H^2 = \Sigma^2 + \frac{1}{3M_P^2}\rho \sim \Sigma^2 +\left(\frac{T^2}{M_P}\right)^2
\,,
\end{equation}
and  by Eq.~(\ref{B-violate}), respectively,  
$T_D$ increases as the shear gets large when $n\geq 1$.
We then find that $Y_B$ becomes larger as larger $T_D$ from Eq.~(\ref{YbIRD}).
The same can be said in the case of the reheating phase.
As a result, we find that the shear also enhances the generated baryon asymmetry 
 through the increase of $T_D$.

Expecting the above two enhancements, we shall 
evaluate the baryon asymmetry in the anisotropic Universe.
If the shear term is dominated comparing with the other terms such that $\rho\ll3M_p^2\Sigma^2$,
we easily find $Y_B$ with such a large shear $\Sigma$.

Here, we focus on the case when the decoupling of the $B$-violating interaction occurs
 during the radiation sub-dominated era\footnote{It is worth mentioning that the gravitational baryogenesis does not  occur in the shear dominated  reheating phase, which is discussed in Appendix \ref{SDR}.}.
Assuming that the shear is dominated, from Eqs.~(\ref{ERad}) and (\ref{A-dotR}), we obtain
\begin{equation}
Y_B \simeq \frac{g_b^2}{2}(1-3w)\frac{T_D^3\Sigma_D}{M_\ast^2M_P^2}. \label{YbAIRD}
\end{equation}
Since the Hubble parameter $H$ is given by 
\begin{equation}
H^2 = \Sigma^2 + \frac{1}{3M_P^2}\rho \sim \Sigma^2,
\end{equation}
the decoupling temperature is evaluated as
\begin{equation}
T_D \simeq \left(\Sigma_D M_B^{2n}\right)^{1/(2n+1)}.
\end{equation}
Substituting this result into Eq.~(\ref{YbAIRD}), the value of $Y_B$ frozen in the 
shear dominated and radiation sub-dominated era is given by
\begin{equation}
Y_B \simeq (1-3w)\Sigma_D^{\frac{2n+4}{2n+1}}M_B^{\frac{6n}{2n+1}}M_\ast^{-2}M_P^{-2}.
\end{equation}

This gives 
 the favorable region of $M_\ast$ and $M_B$ for the anisotropic universe, which
is shown in Fig.~\ref{fig03}.
From this figure, we find that if there exist a large shear during the baryon creation epoch, 
the favorable parameter region is extended, and so the cut-off scale $M_\ast$ can be larger than one in the FLRW case.

\begin{figure}[tbp]
 \centering
 \includegraphics[width=8cm,clip]{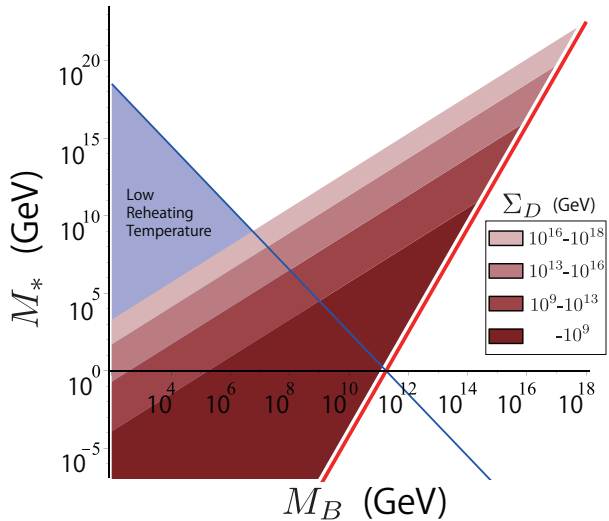}
 \caption{The acceptable range of parameters $M_\ast$ and $M_B$,
  in which 
an observed baryon asymmetry ($Y_B\sim10^{-10}$) is obtained,  in
the case where the decoupling occurs during the shear dominant and radiation sub-dominant era.
We assume a dimension-6 $B$-violating interaction ($n=2$) and 
 the trace anomaly with $1-3w=10^{-1}$.
The pale pink, dark pink, brown, dark red regions are observationally allowed when the shears at the decoupling time are in the ranges of
$10^{18}$ GeV $>\Sigma_D>10^{16}$ GeV, $10^{16}$ GeV $>\Sigma_D>10^{13}$ GeV, $10^{13}$ GeV $>
\Sigma_D>10^{9}$ GeV, $10^{9}$ GeV$ >\Sigma_D$, respectively.
As a reference, the acceptable range for an isotropic universe is shown by the red region. 
When we take into account the gravitino problem ($T_{RD} < 10^9$ GeV) in supersymmetric theories,  
the allowed region is restricted in the shaded region below the blue line.
 }
 \label{fig03} 
\end{figure}

Although this discussion shows the possibility of gravitational baryogenesis in anisotropic 
universe, it has not clarified 
how large the anisotropy can be and whether or not it really 
solves the aforementioned problem in the FLRW spacetime.
It has not given the origin of the anisotropy at all, either.
With all these matters in mind, in this paper, 
we continue to examine whether the anisotropy is really helpful to settle 
the problem of the gravitational baryogenesis
based on a concrete anisotropic inflation model,
 by which we can discuss the origin of anisotropy 
 as well as  its  evolution systematically.

\subsection{Gravitational baryogenesis in anisotropic inflationary model}

 In this paper, we adopt the anisotropic inflationary model proposed in Ref.~\cite{AInf},
 which is shortly summarized in Appendix \ref{Appendix_B}.
The spatial anisotropy is produced during inflation in this model.
Although an anisotropy in the universe will usually disappear during inflation,
  it has been shown that if there exists some coupling with an inflaton 
in the kinetic term of the U(1) gauge field such that 
\bea
-{1\over 4}f(\phi)^2 F_{\mu\nu}F^{\mu\nu}
\,,
\ena
where 
$\phi$ is an inflaton, $F_{\mu\nu}$ is a U(1) gauge field, and the coupling function 
$f(\phi)$ is defined by
\begin{equation}
f(\phi)\equiv\exp\left[{2c\over M_P^2}\int {V\over V'}d\phi \right]
\label{coupling_c}
\end{equation}
with $V(\phi)$ being a potential,
 the anisotropy survives even during inflation and then 
 it can become large at the end of inflation.
Since such a large anisotropy would be important in the anisotropic gravitational 
baryogenesis as well, we reanalyze the evolution of the shear in the anisotropic inflation 
model in Appendix \ref{Appendix_B}.

We assume the model parameter $c$ in (\ref{coupling_c}) 
is smaller than $\mathcal{O}(10)$ 
beyond which the maximum value of $\Sigma/H$ does not change significantly.
The natural value of $c$ will be discussed later.

We denote the energy densities and the pressures of 
radiation fluid, of the inflaton field and of the vector field by  
$\rho_r$, $P_r$,  $\rho_\phi$, $P_\phi$,  and $\rho_v$, $(P_v^x,P_v^y,P_v^z)$,  
which are explicitly given by 
(\ref{DefVE}),  respectively.
Here we have assumed that 
the vector field has the vacuum expectation value in the $x$-direction.
From the symmetry between $y$- and $z$-directions, 
we set $\beta_-=0$ and $\beta_+=\beta$.

Then we  introduce  the total energy density 
$\rho_{\text{tot}}$ and  the EOS parameter $w_{\rm tot}$   by 
\begin{equation}
\rho_{\text{tot}}=\rho_\phi+\rho_v+\rho_r
\,,
\end{equation}
and 
\begin{equation}
w_{\text{tot}} \equiv \frac{P_{\text{tot}}}{\rho_{\text{tot}}} 
= \frac{P_\phi+\bar{P}_v+P_r}{\rho_\phi+\rho_v+\rho_r}
\,,
\end{equation}
respectively, with the average of the pressure of the vector field
\begin{equation}
\bar{P}_v \equiv \frac{1}{3}\left(P_v^x+P_v^y+P_v^z\right)=\frac{1}{3}\rho_v\,,
\end{equation}
and its anisotropic part of the pressure 
\begin{equation}
\Delta P_v \equiv P_v^y-\bar{P}_v=\frac{2}{3}\rho_v
\,.
\label{anisotropic_pressure}
\end{equation}

As mentioned above, the amount of the baryon asymmetry in the gravitational baryogenesis model is proportional to $\dot{R}$.
Since the Ricci scalar in the Bianchi Type I 
Universe is given by
\begin{equation}
R = 6\dot{H}+12H^2+6\Sigma^2\,,
\end{equation}
$\dot{R}$ is evaluated as
\begin{widetext}
\begin{equation}
\dot{R} = -\sqrt{3}(1+w_{\text{tot}})(1-3w_{\text{tot}})\frac{\rho_{\text{tot}} \sqrt{\rho_{\text{tot}}
+3M_P^2\Sigma^2}}{M_P^3} -6(1-3w_{\text{tot}})\frac{\Delta P_v \Sigma}{M_P^2} 
-3\dot{w}_{\text{tot}}\frac{\rho_{\text{tot}}}{M_P^2} 
\label{AI-dotR}
\,,
\end{equation}
\end{widetext}
for the anisotropic inflation model.

Comparing (\ref{AI-dotR}) with the result in the FLRW Universe given by Eq.~(\ref{eqdotR}), 
we find that
the anisotropic component is added to the first term of the right hand side in Eq. (\ref{AI-dotR}) .
Moreover, the second term reflects the effect of the shear evolution by the anisotropic pressure of the vector field.
The last term, as well as Eq.~(\ref{eqdotR}), represents 
the time change of the dominant components.

In order to calculate $\dot{w}$, we have to solve the evolution equation of the matter fields.
This evolution 
depends on the phase where the baryon asymmetry is generated and as in the isotropic case, we consider the following two phases:
(i) reheating phase and (ii) radiation-dominated phase.
We calculate it for each cases and compare the generated baryon asymmetry with that in FLRW Universe.

\subsubsection{Reheating Phase}

Here, we assume that the reheating process occurs only through the perturbative decay 
of inflaton whose rate is expressed by
$\Gamma_\phi$. We adopt 
$\Gamma_\phi /m =10^{-15}$ throughout this paper\footnote{From the standard perturbative reheating, the reheating temperature is given 
by $T_{RD}\sim\sqrt{\Gamma_\phi M_P}$. Thus, we have a constraint  $\Gamma_\phi /m \leq10^{-13}$ in order to obtain sufficiently low 
reheating temperature.}.
It is worth pointing out that this choice  does not change  significantly the final result on the ratio of the generated baryon asymmetry 
as we will see later.
There is also 
an energy transfer between the vector field and the inflaton field,
which transition rate  is given by $\Gamma_v$
We assume that the radiation 
is generated not from the vector field but only from the inflaton field, just for simplicity.
Thus, the energy evolution equations among above three 
matters are given by
\bea
&&\dot{\rho}_v + (4H+4\Sigma)\rho_v = \Gamma_v \rho_v, \label{EVre} \\
&&\dot{\rho}_\phi +3H\rho_\phi = -\Gamma_v\rho_v-\Gamma_\phi\rho_\phi,\label{EIre} \\
&&\dot{\rho}_r +4H\rho_r = \Gamma_\phi\rho_\phi\,.
\label{ERre}
\ena
Using the equation for inflaton field, we find that 
the transition rate  $\Gamma_v$ is described by
\bea
\Gamma_v={2c\over M_P^2}\phi\dot \phi\rho_v
\,.
\ena

Assuming that the EOS parameter for each 
component is constant during this evolution, 
we obtain  $\dot{w}_{\text{tot}}$  from Eqs. (\ref{EVre})-(\ref{ERre}) as
\begin{equation}
\dot{w}_{\text{tot}} = -\frac{H(\rho_v+\rho_r)+4\Sigma \rho_v}{3\rho_{\rm tot}^2} +\frac{\Gamma_v\rho_v+\Gamma_\phi\rho_\phi}{3\rho_{\rm tot}}
\,.
\end{equation} 
Consequently, from Eq. (\ref{AI-dotR}), the time derivative of the Ricci scalar is calculated as
\begin{equation}
\dot{R} = -\sqrt{3}\frac{\rho_\phi\sqrt{\rho_{\text{tot}}+3M_P^2\Sigma^2}}{M_P^3} -\frac{\Gamma_v\rho_v+\Gamma_\phi\rho_\phi}{M_P^2} 
\,.
\end{equation}
Since $\Gamma_v$  is decreasing in the reheating process, thus we may neglect the term including $\Gamma_v$ in this phase.
As a result, we obtain 
\begin{equation}
\dot{R} = -\sqrt{3}\frac{\rho_\phi\sqrt{\rho_{\text{tot}}+3M_P^2\Sigma^2}}{M_P^3} -\frac{\Gamma_\phi\rho_\phi}{M_P^2} 
\label{dotR_BI}
\,.
\end{equation}

In the standard FLRW Universe without the vector field,
we find 
\begin{equation}
\dot{R} = -\sqrt{3}\frac{\rho_\phi\sqrt{\rho_{\text{tot}}}}{M_P^3} -\frac{\Gamma_\phi\rho_\phi}{M_P^2}
\label{dotR_FLRW}
\,.
\end{equation}
The difference between (\ref{dotR_BI}) and (\ref{dotR_FLRW}) 
 is only the part inside the square root,  which is just given by
  the Hubble parameter.
Since the Hubble parameter becomes large by the existence of the shear, 
we expect 
the generated baryon asymmetry is consequently enhanced in the anisotropic model.

We start to calculate the baryon asymmetry from the end of inflation 
$t=t_e$ for the anisotropic inflation model with the potential 
$V(\phi)={1\over 2}m^2\phi^2$.
We then compare the amount of the generated baryon asymmetry with 
that in the FLRW Universe with the same set-up, i.e., 
 we assume that the value of  $M_B$ is the same, 
 which means that we fix the same $B$-violating interaction\footnote{
Here, we have assumed the $B$-violating interaction is given by a mass dimension 6 ($n=2$) operator.
},
and that the inflaton mass $m$ and the definition of the end of inflation are the 
same in both models.
In Fig.~\ref{ReDifYB}, we 
plot the ratio of the baryon asymmetry in the anisotropic model $Y_{B,\,\text{aniso}}$ 
and that in the standard isotropic model $Y_{B,\,\text{iso}}$ 

\begin{figure}[h]
 \centering
 \includegraphics[width=80mm]{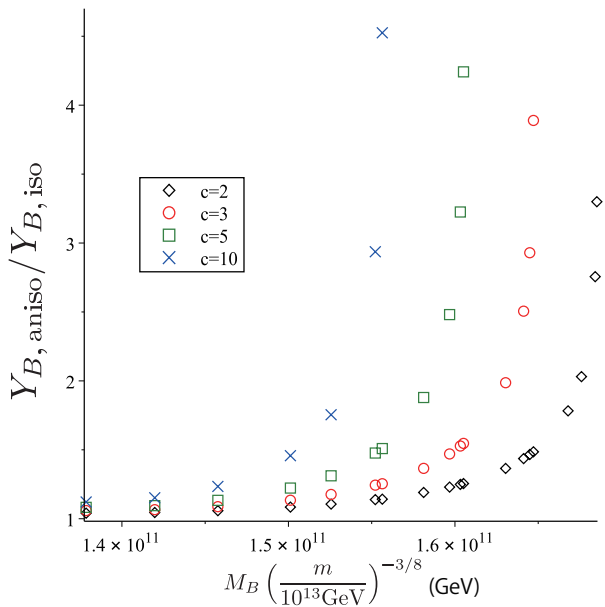}
 \caption{
The difference of the generated baryon asymmetry between the anisotropic/isotropic 
models are plotted in terms  of  $M_B$.
We assume that the $B$-violating interaction ($n=2$)
is decoupled in the reheating phase.
We choose the initial data from the numerical calculation of anisotropic inflation at $t=t_e$.
and assume $\Gamma_\phi /m =10^{-15}$.
 The black (diamond), red (circle), green (box), and cyan (cross) plots represent the results for the cases of $c=2, 3, 5,$ and $
  10$, respectively.
}
 \label{ReDifYB}
\end{figure}

We must mention that we have used the following assumptions in the above calculations.
As we showed in Sec.~\ref{GBGandEx}, the additional entropy creation during the reheating phase dilutes the generated baryon asymmetry.
This dilution effect can be treated easily in our calculation with the following assumption.
Since we are focusing our attention on the case that the decoupling of the $B$-violating interaction occurs during the reheating phase, the reheating must be proceeded slowly.
On the other hand, the shear and the vector energy density is diluted faster than the inflaton energy density, and so the Universe immediately comes to resemble with FLRW Universe.
Hence we assume that  the dilution factor is the same as one of 
the FLRW Universe case, i.e., $(T_{RD}/T_D)^5$.
Besides, we know that the gradual reheating makes the reheating temperature depending only on the interaction rate $\Gamma_\phi$.
This is because the reheating process  will
finish at $\Gamma_\phi\simeq H$ and the temperature is 
given by the energy density of the radiation which satisfies
\bea
H^2 = \Sigma^2 +\frac{1}{3M_P^2}\rho_{\text{tot}} \simeq \frac{1}{3M_P^2}\rho_r,
\ena
where we have used the fact that the shear and the vector energy density are diluted faster than the inflaton energy density.
Thus the reheating temperature $T_{RD}$ are assumed to be the same in both models, and so the difference of the dilution factor 
depends only on the difference of the decoupling temperature $T_D$.

From Fig.~\ref{ReDifYB}, we find the enhancement effect of 
generating baryon asymmetry by the effect of the anisotropic component.
Since, as mentioned in Appendix \ref{Appendix_B}, the generated anisotropy of the universe increases as the parameter $c$ 
gets large, 
we find that the amount of the baryon asymmetry becomes larger for the larger value of $c$.

If we do not take into account the constraint  from the gravitino problem, 
$Y_{B,\,\text{aniso}}$  can be larger by one order of magnitude than $Y_{B,\,\text{iso}}$.
However, if we impose such a constraint, 
the enhancement factor is larger just by a few times,
which  may not be sufficient.
Since  $Y_B\propto M_\ast^{-2}$, 
the enhancement factor  does not allow the observationally 
favorable value of $M_\ast$ which must be sufficiently higher than the LHC scale.
We conclude
 that it is difficult to obtain a sufficient baryon asymmetry
 in the reheating phase.

In addition to this, as the shear decreases faster than the other fields, its enhancement effect  disappears rapidly.
We find from Fig.~\ref{ReDifYB} that the effect of anisotropic components appear 
in the short range of $M_B$, which value is related 
to the decoupling temperature.
Therefore, the baryon asymmetry is affected by the anisotropic effect 
only in the case that the $B$-violating interaction is 
decoupled immediately after the end of inflation. 
In the other words, the fine-tuning of model parameters is required
in order for the anisotropic components to affect the generated baryon asymmetry.

In the above calculation, we have assumed that the reheating process starts at $t=t_e$.
However, there is an ambiguity when the reheating really starts.
Hence, we have also checked how 
 the enhancement factor $Y_{B,\,\text{aniso}}/Y_{B,\,\text{iso}}$ 
is sensitive to 
the starting time of the reheating 
 by  choosing the latest possible staring time $t=t_f$ which is 
 when the inflaton field reaches $\phi=0$.
We find that the result does not depend on the starting time of the reheating so much (within the factor of 1$\pm$0.2). 

Since we assume that the gradual reheating, 
as long as $\Gamma_\phi$ is sufficiently small, the anisotropic components 
completely diluted before a certain amount of  radiation energy density appears in the total energy density.
With such a small value of $\Gamma_\phi$, the enhanced baryon asymmetry does not much change 
for the different values of $\Gamma_\phi$, as the backreaction of radiation is neglected in the initial phase of reheating.
Therefore, the different choice of $\Gamma_\phi$ affects only the temperature of the Universe.
It means that the fine-tuning of $M_B$ depends strongly on the value of $\Gamma_\phi$.

Finally, we comment about  the 
 mass dimension of $B$-violating interaction.
 The case of the mass dimension 6 ($n=2$), which we have analyzed in this paper, may gives the maximum enhancement.
 If the mass dimension is 5, the decoupling of $B$-violating interaction never happens because
  $\Gamma_B$ does not decrease
   faster than the Hubble expansion rate $H$.
For the case of the mass dimension higher than 6 ($n=2$), 
we expect that 
the increase of $T_D$ by the shear becomes inefficient, 
and it diminishes the generated baryon asymmetry.

\subsubsection{Radiation-dominated Phase}

Next we evaluate $\dot{R}$ in the radiation dominated phase.
In this phase, there is no longer the interaction between the inflaton field and the vector or the radiation field.
Therefore, the evolution equations of the energy densities are given by
\bea
&&\dot{\rho}_v + (4H+4\Sigma)\rho_v = 0, \label{EVrd}\\
&&\dot{\rho}_r +(4-3\varepsilon)H\rho_r = 0\,,\label{ERrd}
\ena
where $\varepsilon$ is the trace anomaly, which magnitude is expected to be  $10^{-2}$-$10^{-1}$.
The effective EOS parameter is expressed by
\begin{equation}
w_{\text{tot}} = \frac{\bar{P}_v+P_r}{\rho_v+\rho_r} = \frac{1}{3}\frac{\rho_v+(1-3\varepsilon)
\rho_r}{\rho_v+\rho_r}.
\end{equation}
Thereby, from the evolution equations (\ref{EVrd}) and (\ref{ERrd}), $\dot{w}_{\text{tot}}$ is solved up to second order of $\varepsilon$ as
\begin{equation}
\dot{w}_{\text{tot}} = 
-\frac{\varepsilon  (4\Sigma +3\varepsilon H)\rho_v\rho_r}{(\rho_r +\rho_v)^2}\,.
\end{equation}
Consequently, the time derivative of the Ricci scalar becomes
\begin{equation}
\dot{R} = -\frac{\varepsilon(4-3\varepsilon) }{M_P^3}\rho_r \sqrt{3(\rho_{\text{tot}}+3M_P^2\Sigma^2)}
\,.
\end{equation}

In the standard FLRW Universe,
since  the component of the matter is only the radiation,
we  find
\begin{equation}
\dot{R} = 
-\frac{\varepsilon(4-3\varepsilon) }{M_P^3}\sqrt{3}\rho_r^{3/2} \,.
\end{equation}

\begin{figure}[thbp]
 \centering
 \includegraphics[width=80mm]{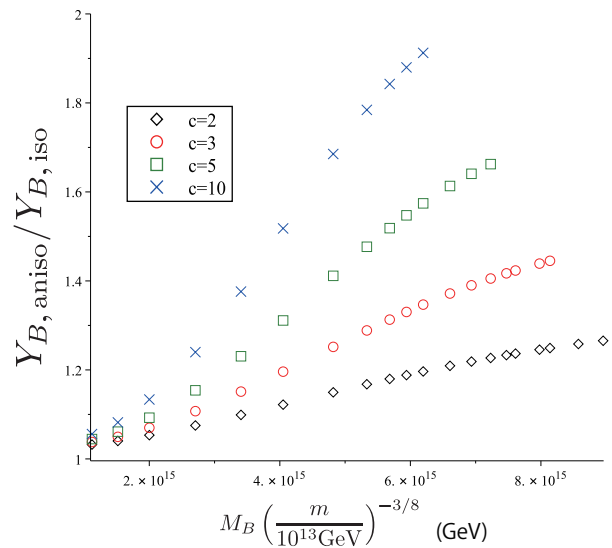}
 \caption{
The difference of generated baryon asymmetry between anisotropic/isotropic
models, plotted as a function of $M_B$.
In this picture, we assumed that the $B$-violating interaction ($n=2$)
is decoupled in the radiation dominated phase.
Additionally, we choose the initial data from the numerical calculation of anisotropic inflation at $t=t_e$.
Also, we assume trace anomaly is $\varepsilon=10^{-1}$.
The black (diamond), red (circle), green (box), and cyan (cross) plots represent the results for the cases of $c=2, 3, 5, $ and 
$10$, respectively.
}
 \label{RdDifYB}
\end{figure}

If the reheating takes for long time, 
just  as the previous reheating case, the anisotropic component is
completely diluted and it will not affect the generation of the baryon asymmetry.
Therefore, in this case, we assume that the reheating process rapidly finishes.
Assuming that the instantaneous reheating occurs, the initial radiation energy density 
is given by  the inflaton energy density.
We show the generated baryon asymmetry  in Fig.~\ref{RdDifYB}.

As the previous reheating case, we find that the enhancement factor
 is at most  $\mathcal{O}(1)$.
We find that the ambiguity of when the reheating really starts
does not affect the result significantly (within the factor of 4).

We find that the enhancement factor $Y_{B,\;\text{aniso}}/Y_{B,\;\text{iso}}$ is always 
smaller than that in the reheating case.
We also find that  the ratio is less sensitive to $M_B$ 
 compared with one in the reheating case.
Thus, the anisotropic component contribution 
 is less efficient in the radiation dominated phase, 
 but the tuning for $M_B$ is not required strongly.

Furthermore, when we take into account the gravitino problem, 
the expected reheating temperature and the existence of the anisotropic components in the 
radiation-dominated phase may not be compatible.
Since the energy scale of the inflation seems much higher than the reheating temperature constrained by the gravitino problem, the assumption of an instantaneous reheating process is contrary to this low reheating temperature.
Therefore, we conclude that it is difficult to consider a sufficiently large initial anisotropy in the radiation dominated phase.
As a result,  the anisotropic components generated in the anisotropic inflation do not improve the gravitational baryogenesis enough.

\subsection{Observational Constraint on $c$}

In the above calculation, we have assumed the model parameter $c$ as a free parameter.
However, $c$ is strongly constrained by the observation of CMB anisotropy.

The statistical anisotropy in the power spectrum of the curvature perturbation is parametrized by \cite{CMBAdif}
\begin{equation}
P(\bold{k}) = P(k)\left[1+h_\ast (\hat{\bold{k}}\cdot\hat{\bold{v}})\right],
\end{equation}
where $\hat{\bold{v}}$ is some preferred direction in space and $h_\ast$ is the amplitude of the anisotropy.
As shown by \cite{CMBA}, $h_\ast$ is calculated 
for the anisotropic power law inflation as 
\begin{equation}
h_\ast = 24\left(1-\frac{1}{c}\right)N_k^2,
\end{equation}
where $N_k$ is the number of $e$-folds of the fluctuation with the wave number $k$ 
 counted from the end of the inflation.
The current observational bound on $h_\ast$ is given by 
 $0.002\pm0.016$ (68\% CL) \cite{obsA}.
The model parameter $c$ is strongly bounded as
\begin{equation}
1-\frac{1}{c} \lesssim 10^{-7}\times \left(\frac{|h_\ast|}{10^{-2}}\right)\left(\frac{N_k}{60}\right)^{-2}.
\end{equation}
Thus we find that $c$ must be extremely close to unity, so that the anisotropic component 
does not  grow up so largely.
It is difficult to expect a sufficiently large anisotropy of the Universe from the anisotropic inflation
with satisfying the observational constraints. 

Note that the above calculation is based on the assumption that the model parameter $c$ satisfies $c-1$$\gtrsim\mathcal{O}(1)$ 
so that the attractor solution of the anisotropic inflation exists. 
Using another branch of the anisotropic inflation
found in Ref.~\cite{CMBreA}, 
it was shown that the allowed parameter region is slightly increased.

\section{Conclusion and Discussion}

We have discussed the gravitational baryogenesis mechanism in the anisotropic spacetime
 induced by the anisotropic inflation.
The main purpose of the present work has been to check a possibility of 
the gravitational baryogenesis
by the anisotropic-shear enhancement.
We have obtained the following results using the concrete anisotropic inflationary model.

First, we have clarified the issue of gravitational baryogenesis that is unavoidable if the temperature of the universe is bounded by the gravitino problem.
The cut-off mass parameter $M_\ast$ is smaller than the LHC scale in order to generate a sufficient baryon asymmetry.
This unnaturalness denies a possibility of baryogenesis via gravitational interactions.
However it might be resolved if the background spacetime is anisotropic.
This is due to the shear which enhances 
$\dot{R}$ and $T_D$, and so
the amount of generating baryon asymmetry increases.
For instance, if there exists a huge shear such as $\Sigma_D\gtrsim10^9$ GeV 
in the radiation dominated phase, the favorable $M_\ast$ is enlarged above the LHC scale.
Thus we find that  there is a little possibility of  gravitational baryogenesis.

Next, we have studied a concrete example of the anisotropic spacetime 
caused by the anisotropic inflation.
 Although the shear is negligible during the inflation, it can grow 
 exponentially at the end of the inflation.
Moreover, we have seen that the anisotropy of the universe $\Sigma/H$ can increase 
to $O(1)$ at the end of the inflation if the model parameter $c$ takes sufficiently large value.
We find that  the anisotropy works to increase the baryon asymmetry,
but the enhancement factor is not large.
Furthermore, as the anisotropic components are rapidly diluted comparing with the other isotropic mater fields after the inflationary phase is finished, the enhancement works only in a brief period.
In other words, the anisotropic effect appears if and only if $B$-violating interaction decouple is occurred immediately after the end of inflation.
Thus, we newly suffer from the fine-tuning problem for $M_B$. 
The most crucial problem is that $c$ must be extremely close to unity from the observational constraint by the statistical anisotropy in the power spectrum of the curvature perturbation.
Due to this, it is hopeless to generate large anisotropy of the universe by the anisotropic inflationary model.
The model of gravitational baryogenesis has been still suffered from a low temperature bound by the gravitino problem.

It should be noted that in this paper we have assumed the anisotropy of the universe is 
originated by the anisotropic inflationary model.
If there exists another mechanism for generating huge anisotropy,
 gravitational baryogenesis might be reactivated again.

Finally, we comment on the reheating temperature.
In this paper, we have used the standard constraint for the reheating temperature
 by the gravitino problem.
This constraint, however,  might be relaxed in some supergravity models.
In the case of the heavy gravitino mass, it will decay before the BBN epoch 
but the overproduced LSPs overclose the Universe.
However, if the $R$-parity is violated and LSPs can decay before the 
BBN epoch, we can avoid the overproduction of LSPs.
Then the high reheating temperature can be arrowed with such a 
$R$-parity violating interaction.

\section*{Acknowledgments}

We would like to thank Kohei Kamada and Fuminori Hasegawa
for useful discussions and comments.
 This work was supported in part by Grants-in-Aid from the 
Scientific Research Fund of the Japan Society for the Promotion of Science 
(No. 25400276 and No. 26887042).


\newpage


\appendix
\section{Shear dominated Reheating phase}\label{SDR}
\label{Appendix_A}

Using the Bianchi Type I metric, Friedmann equation is given by
\begin{equation}
H^2 = \Sigma^2 +\frac{1}{3M_P^2}\rho,
\end{equation}
where $\Sigma$ and $\rho$ are the shear and energy density of the matter fields, respectively.
Here, we assume that the case that the shear is dominated comparing with the other matter fields; $M_P^2\Sigma^2 \gg \rho$.
On the other hand, we know that the shear evolves as $\Sigma\propto a^{-3}$, where $a$ is the spatially averaged scale factor.
Therefore, from the Friedmann equation, the Universe is expanded as $a\propto t^{1/3}$ in the shear dominated epoch.

With this in mind, we discuss the evolution of the radiation energy density 
in the reheating phase. 
The radiation energy density 
is given by Eq. (\ref{eq_reh_rad}).
If the reheating processes gradually proceed ($\Gamma_\phi t_{\text{osc}} \ll 1$), Eq. (\ref{eq_reh_rad}) can be solved as
\begin{equation}
\rho_r (t) \simeq \frac{3}{4}M_I^4 \Gamma_\phi t_{\text{osc}} \left[1-\left(\frac{t_{\text{osc}}}{t}\right)^{4/3}\right],
\end{equation}
where we have used $a\propto t^{1/3}$ in the shear dominated case.
Thus, the leading term of the radiation energy density is constant in this phase.

This result means that the amount of the radiation energy density generated by the inflaton field is almost balanced with its dilution effect by the cosmic expansion.
That is why the temperature can be regard constant during the shear dominated reheating phase.
By contrast, the gravitational baryogenesis needs a decoupling of the $B$-violating interaction. 
We, however, cannot expect it, if the background temperature is constant.
That is because the constant temperature makes $\Gamma_B$ be also constant with time, on the other hand $H\sim\Sigma$ monotonically decreases in this phase.
Accordingly, we conclude that the gravitational baryogenesis does not occur during the shear dominated reheating epoch.

\section{Anisotropic Inflation}
\label{Appendix_B}
In this Appendix, we shortly summarize the anisotropic inflationary model
proposed in \cite{AInf}.

To be precise, we consider the following action:
\begin{widetext}
\begin{equation}
\mathcal{S} = \int d^4x \sqrt{-g}\left[\frac{M_P^2}{2} R 
-\frac{1}{2}(\partial_\mu\phi)(\partial^\mu\phi) - V(\phi)
-\frac{1}{4}f(\phi)^2F_{\mu\nu}F^{\mu\nu}\right], \label{AIact}
\end{equation}
\end{widetext}
where 
$\phi$ is an inflaton field, 
$F_{\mu\nu}$ is the field strength of a U(1) gauge field 
defined by $F_{\mu\nu} = \partial_\mu A_\nu-\partial_\nu A_\mu$
with the vector potential $A_\mu$,
and $V(\phi)$ is the inflaton potential.
The coupling function
$f(\phi)$ is assumed to be 
\begin{equation}
f(\phi)=\exp\left[\frac{2c}{M_P^2}\int\frac{V}{V'}d\phi\right], \label{DifF}
\end{equation}
where the parameter $c$ describes the strength of the coupling to the gauge field and 
a prime denotes a derivative with respect to $\phi$.
For a chaotic inflation with $V(\phi)={1\over 2}m^2\phi^2$, we 
find $f(\phi)=\exp\left[\frac{c\phi^2}{2M_P^2}\right]$.
It is shown that the anisotropic inflation is realized if $c>1$~\cite{AInf}. 
 
For the gauge field, we assume
\begin{equation}
A_\mu dx^\mu = v(t) dx\,,
\end{equation} 
which guarantees a homogeneous universe.
Since there exists a rotational symmetry in the $y$-$z$ plane, 
 we find an axially symmetric Bianchi Type I geometry, which metric is 
 given by
\begin{equation}
ds^2=-dt^2+e^{2\alpha(t)}\left[e^{-4\beta (t)}dx^2
+e^{2\beta(t)}\left(dy^2+dz^2\right)\right], \label{MAnsatz}
\end{equation}
where $t$ is a  cosmic time, and
 $e^{3\alpha}$ and $\beta$ denote a three-volume and a spatial anisotropic metric variable, respectively.
Under this ansatz, we can solve the equation of motion for the vector field as
\begin{equation}
\dot{v}=f^{-2}e^{-\alpha-4\beta} C_A, \label{SolVec}
\end{equation}
where $C_A$ is a constant of integration.

The energy-momentum tensors of the inflaton and the vector field are given by
\bea
&&T_{\mu}{}^{\nu}(\phi)=\left(-\rho_\phi, P_\phi, P_\phi, P_\phi\right) ,\\
&&T_{\mu}{}^{\nu}(v)=\left(-\rho_v, P_v^x, P_v^y, P_v^z\right),
\ena
where 
\bea
&&\rho_\phi=\frac{1}{2}\dot{\phi}^2+V
\,,
\nn
&&P_\phi= \frac{1}{2}\dot{\phi}^2-V\,,
\nn
&&\rho_v = \frac{C_A^2}{2}f^{-2}e^{-4\alpha-4\beta} \,,
\nn
&&P_v^x = -\frac{C_A^2}{2}f^{-2}e^{-4\alpha-4\beta} = -\rho_v \,,
\nn
&&P_v^y=P_v^z = \frac{C_A^2}{2}f^{-2}e^{-4\alpha-4\beta} = \rho_v
\,.
\label{DefVE}
\ena

We define the average of the pressure of the vector field by
\begin{equation}
\bar{P}_v \equiv \frac{1}{3}\left(P_v^x+P_v^y+P_v^z\right)=\frac{1}{3}\rho_v\,,
\end{equation}
and
the variation of the pressure $\Delta P_v$ 
by 
\begin{equation}
\Delta P_v = \frac{2}{3}\rho_v
\,,
\label{anisotropic_pressure}
\end{equation}
i.e.,
\begin{numcases}
{ }
P_v^x = \bar{P}_v -2\Delta P_v, \\
P_v^y = P_v^z = \bar{P}_v +\Delta P_v\,.
\end{numcases}
We also use the EOS parameter of the vector field in each direction by
\bea
&&w_v^x \equiv \frac{P_v^x}{\rho_v} = -1,\\
&&w_v^y =w_v^z \equiv \frac{P_v^y}{\rho_v} = 1,\\
&&w_v \equiv \frac{\bar{P}_v}{\rho_v} = \frac{1}{3}.
\ena
The spatially averaged vector field behaves as the relativistic particles.

The basic equations are given by
\bea
&&\dot{\alpha}^2 = \dot{\beta}^2 + \frac{1}{3M_P^2}\left[\rho_\phi+\rho_v\right] ,  \label{Beq1} \\
&&\ddot{\alpha}=-3\dot{\alpha}^2+\frac{1}{M_P^2} V +\frac{\rho_v}{3} \label{Beq2}, \\
&&\ddot{\beta}=-3\dot{\alpha}\dot{\beta} +\frac{2}{3M_P^2}\rho_v \label{Beq3}, \\
&&\ddot{\phi}=-3\dot{\alpha}\dot{\phi}-V'+2 f^{-1}f' \rho_v\label{Beq4},
\ena
where
\bea
\rho_\phi\equiv \frac{1}{2}\dot{\phi}^2+V
\,,~~
\rho_v\equiv\frac{C_A^2}{2}f^{-2}e^{-4\alpha-4\beta}
\,.
\ena

\begin{figure}[h]
 \centering
 \includegraphics[width=80mm]{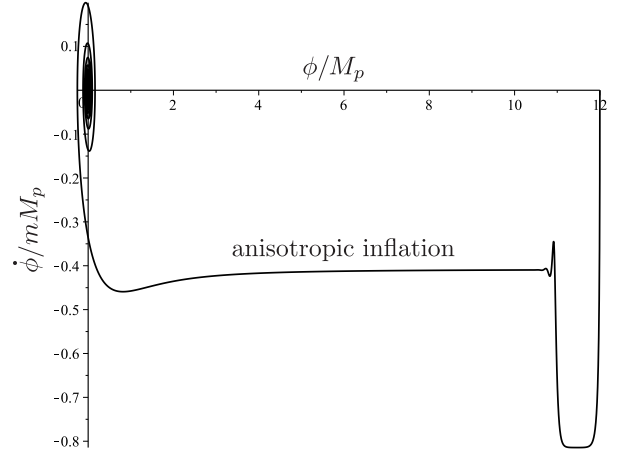}
 \caption{
Phase diagram of the evolution of the inflaton field in the anisotropic inflation model.
The curve shows the evolution path in the $\phi$-$\dot{\phi}$ phase space for the case with $c=2$.
The transition between the first isotropic and the second anisotropic inflations occur around $\phi/M_P=11$.
After the transition, $\dot{\phi}$ becomes slower by $1/c$ times than that in the previous phase.
The inflaton field eventually enters into the oscillating phase after the end of inflation.
}
 \label{phi_Dphi}
\end{figure}

Adopting the inflaton potential 
\begin{equation}
V(\phi) = \frac{1}{2}m^2\phi^2, \label{InfPot}
\end{equation}
where $m$ is the inflaton mass, 
we have numerically studied the anisotropic inflationary model, 
and analyzed the evolution of the shear in detail.
The gauge kinetic function  is now
\begin{equation}
f(\phi)=e^{\frac{c}{2M_P^2}\phi^2}.
\end{equation}

This inflationary model possesses the following two phases:
 (i) the conventional slow-roll inflationary phase and (ii) the anisotropic inflation after 
the contribution of the vector field 
becomes no longer negligible.
If the initial energy density of the vector field is much smaller than that of the inflaton,
the first phase is realized.
During this phase,
since the energy density of the vector field grows as $\rho_v\propto e^{4(c-1)\alpha}$, 
the vector field eventually comes
to affect the dynamics of the inflaton field and the second anisotropic inflationary 
phase appears\footnote{However, if there exist three or more U(1) fields with the same coupling function or the Yang-Mills field with the similar coupling to the inflaton field, 
an isotropic inflationary expansion
becomes an attractor\cite{inflation_w_gauge, inflation_w_YM}.}.
We show one example in Fig. \ref{phi_Dphi}.

During the inflationary phase,
we find that the effect of shear cannot 
become as large as the Hubble parameter.
The measure of the anisotropy is described by $\Sigma/H$, where $H\equiv\dot{\alpha}$ 
and $\Sigma\equiv |\dot{\beta}| $ describe the average expansion rate and
 the magnitude of the spacetime shear, respectively.
During the anisotropic inflation when
the slow-roll approximation is valid, it is shown that the above anisotropic parameter satisfies \cite{AInf}
\begin{equation}
\frac{\Sigma}{H}  = \frac{1}{3}\frac{c-1}{c}\epsilon_H, \label{S/H}
\end{equation}
where 
the slow-roll parameter 
in terms of the Hubble parameter is defined by
\begin{equation}
\epsilon_H \equiv -\frac{\dot{H}}{H^2}.
\end{equation}
Eq. (\ref{S/H}) means that $\Sigma/H$ is approximately the same as 
the slow-roll parameter $\epsilon_H$, unless  $|c-1|\ll 1$.
Therefore, in spite of the
strong coupling $c$, the anisotropic parameter $\Sigma/H$ is much smaller than unity in the inflationary stage.

\begin{figure}[h]
 \centering
 \includegraphics[width=80mm]{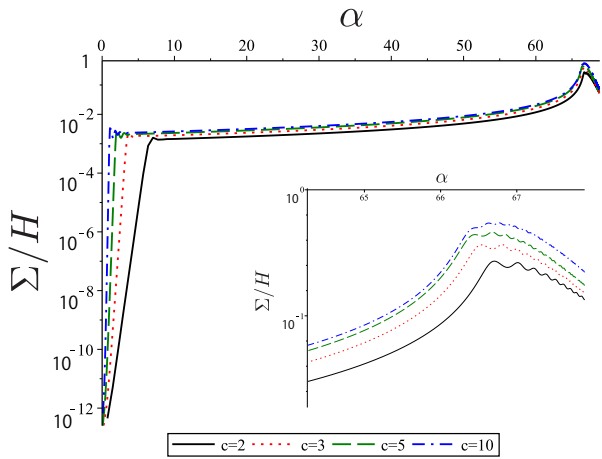}
 \caption{
The evolution of the anisotropy  parameter $\Sigma/H$ with respect to the $e$-foldings.
In the first isotropic inflationary phase, the anisotropy grows rapidly up to $\mathcal{O}(\epsilon_H)$ 
in accordance with the increment of the vector energy density.
While it becomes almost constant 
 during the second anisotropic inflationary phase.
Then it suddenly grows again just at the end of the inflation.
The black (solid), red (dotted), green (dashed), and 
blue (chain) plots show the cases of $c=2, 3, 5, 10$, respectively.
We also enlarge the figure near the peaks to see the dependence of 
$c$ more clearly.
}
 \label{AI_a}
\end{figure}

However,  as shown in Fig. \ref{AI_a}, 
the shear increases exponentially at the end of inflation.

In the evaluation of the generated baryon asymmetry in the text, 
we have used our numerical solution,
but the increase of the shear at the end of inflation can be 
obtained (or restricted) by the semi-analytic approach, which we will show 
 in the next section.

\section{Evaluation of the shear in anisotropic inflation}

In this Appendix, we derive an upper limit of the spacetime shear 
at the end of the anisotropic inflation by the semi-analytic 
approach (See also \cite{AInf_max-S}).

First we see the existence of the saturation of the shear magnitude 
from our numerical calculation.
Using Eqs.~(\ref{Beq1}) and (\ref{Beq2}), 
we describe the shear in terms of the slow-roll parameter $\epsilon_H$ as
\begin{equation}
\Sigma^2  = \frac{(2\epsilon_H-3-3w_{\text{tot}}) \rho_{\text{tot}}}{6M_P^2 (3-\epsilon_H)}\,.
\label{Sigma_ito_epsilon}
\end{equation}
Since the right hand side of Eq.~(\ref{Sigma_ito_epsilon}) includes $\rho_{\text{tot}}$,
 its dependence on $\epsilon_H$ seems complicated.
During the inflationary era satisfying $\epsilon_H \leq 1$, however, 
  we find that $\Sigma$ increases monotonically as $\epsilon_H$ increases
from the numerical calculation as shown in Fig.~\ref{shear_SR}.
The upper limit of the shear 
seems to exist and its value is evaluated 
at $\epsilon_H \approx 1$, which is
\begin{equation}
\Sigma^2 =  -\frac{1}{12M_P^2}(1+3w_{\text{tot}})\rho_{\text{tot}}\,.
\end{equation}

From Eq.~(\ref{Beq1}), the anisotropy $\Sigma/H$ has 
the upper bound as
\begin{equation}
\frac{\Sigma}{H}\Big{|}_{\rm max}
= \sqrt{-\frac{1+3w_{\text{tot}}}{3(1-w_{\text{tot}})}}. \label{SHatE}
\end{equation}
If we impose the weak energy condition on 
the effective matter field, $w_{\text{tot}}$ must be larger than $-1$.
As a result, the anisotropy must satisfy 
\begin{equation}
0 < \frac{\Sigma}{H} < \frac{1}{\sqrt{3}}.
\end{equation}
This  is the upper bound of the 
spatial anisotropy in Bianchi Type I Universe 
during the inflationary era.

\begin{figure}[h]
\vskip 1cm
 \centering
 \includegraphics[width=60mm]{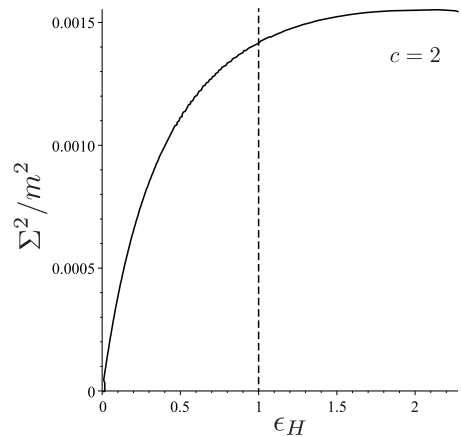}
 \caption{
Plots of the shear $\Sigma^2$ vs slow-roll parameter $\epsilon_H$ with $c=2$.
The vertical broken line expresses $\epsilon_H=1$ .
The upper bound of the shear is evaluated at the intersection point.
}
 \label{shear_SR}
\end{figure}

We then evaluate the more precise upper bound of anisotropy 
by the semi-analytic approach.
For a chaotic inflation with $V(\phi)=\frac{1}{2}m^2\phi^2$, $H$ and $\phi$ can be approximated 
by linear functions of the cosmic time during the second inflationary phase:
\begin{align}
H(t)&=-\frac{m^2}{3c}(t-t_t)+H(t_t) \, , \\
\phi(t)&=-\frac{\sqrt{6}mM_P}{3c}(t-t_t)+\frac{\sqrt{6}M_P}{m}H(t_t)\, ,
\end{align}
where $t_t$ denotes the transition time from first isotropic 
inflationary phase to the second anisotropic inflationary phase.
We have also used the approximation that the energy density of the vector field is given by the following constant
\begin{equation}
\rho_v=\frac{c-1}{2c^2}m^2M_P^2\,, \label{VE-phi^2}
\end{equation}
during anisotropic inflation \cite{AInf}.
Substituting these into the evolution equation of the shear 
given by Eq.~(\ref{Beq3}), we find
\begin{widetext}
\begin{eqnarray}
\frac{\Sigma(t)}{m} &=& A(t)
\exp\left[\frac{m^2}{2c}\left(t-t_t-\frac{3c}{m^2}H(t_t)\right)^2\right]\,, \label{AnaS1}
\label{Sigma_max}
\end{eqnarray}
with 
\begin{eqnarray}
A(t)&=& \left[\frac{\sqrt{2\pi}}{6}\frac{c-1}{c^{3/2}}\left\{\text{erf}\left(\frac{3\sqrt{c}H(t_t)}{\sqrt{2}m}\right) -\text{erf}\left(-\frac{m}{\sqrt{2c}}(t-t_t)+\frac{3\sqrt{c}H(t_t)}{\sqrt{2}m}\right) \right\}+\frac{\Sigma(t_t)}{m}\exp\left(-\frac{9cH(t_t)^2}{2m^2}\right)\right] 
\,, \label{AnaS2}
\end{eqnarray}
\end{widetext}
where $\text{erf}(x)$ denotes the error function.

The anisotropic inflation takes place between $t=t_t$ and $t_e$, where 
\begin{equation}
t_e=t_t+\frac{3c}{m^2}H(t_t)-\frac{\sqrt{3c}}{m}\,.
\end{equation}
is the end time of inflation evaluated by $\epsilon_H=1$.
Although the exponential function in Eq. (\ref{Sigma_max}) does not 
change so much during the above inflationary period,
 the amplitude $A(t)$ grows rapidly and saturates around 
the end of inflation, just because of the 
typical behavior of the error function.
As a result, we find that the 
shear increases drastically during the anisotropic inflationary phase.

We expect from the above saturation of $\Sigma$ 
that the anisotropy of the Universe becomes 
the largest at the end of the inflation.
The maximal value  is obtained at $t=t_e$ as

\begin{equation}
\frac{\Sigma}{H}\simeq\sqrt{\frac{\pi e^3}{6}}\left[1-\text{erf}\left(\sqrt{\frac{3}{2}}\right)\right] \left(1-\frac{1}{c}\right)\,. \label{Ana_AI}
\end{equation}
Here, we have used $cH(t_t)^2/m^2\gg1$.

This result gives the better evaluation for the maximal value of anisotropy comparing with the previous result Eq.~(\ref{S/H}).
The other approach to evaluate the maximal value 
by use of the higher-order expansion of the slow-roll parameter 
has been also given \cite{AInf_max-S}.
Our result is mostly the same as theirs.

Eq.~(\ref{Ana_AI}) gives a good explanation for the tendency that 
the anisotropy of the Universe at the end of inflation is larger
as the parameter $c$ increases as well as the fact that it is saturated for sufficiently large $c$.
Therefore, we do not expect an enormously large anisotropy 
at the end of the anisotropic inflation even with the 
very large $c$.



\end{document}